\documentclass[notitlepage,aps,prl,reprint,twocolumn,longbibliography,superscriptaddress]{revtex4-1}
\usepackage{graphicx}
\usepackage{amsmath}
\usepackage{amssymb}
\usepackage{comment}
\usepackage[colorlinks, allcolors=blue]{hyperref}
\usepackage[all]{hypcap}
\usepackage[mathlines]{lineno}
\usepackage{braket}
\usepackage{wrapfig}
\usepackage{lipsum}
\usepackage{ulem}
\usepackage{xspace}

\newcommand{\pref}[2]{\hyperref[#1]{\ref{#1}(#2)}}
\newcommand{\preff}[2]{\hyperref[#1]{\ref{#1}#2}}
\newcommand{\eqpref}[1]{\hyperref[#1]{(\ref{#1})}}

\newcommand{\squig}{{\raise.17ex\hbox{$\scriptstyle\sim$}}}

\newcommand{\aub}{Aubry-Andr\'{e}\xspace}

\begin{document}
\title{Observation of tunable mobility edges in generalized Aubry-Andr\'{e} lattices}
\author{Fangzhao Alex An}
\thanks{These authors contributed equally to this work.}
\affiliation{Department of Physics, University of Illinois at Urbana-Champaign, Urbana, IL 61801-3080, USA}
\author{Karmela Padavi\'{c}}
\thanks{These authors contributed equally to this work.}
\affiliation{Department of Physics, University of Illinois at Urbana-Champaign, Urbana, IL 61801-3080, USA}
\author{Eric J.~Meier}
\affiliation{Department of Physics, University of Illinois at Urbana-Champaign, Urbana, IL 61801-3080, USA}
\author{Suraj Hegde}
\affiliation{Max-Planck Institute for Physics of Complex Systems, 01187 Dresden, Germany}
\author{Sriram Ganeshan}
\email{sganeshan@ccny.cuny.edu}
\affiliation{Physics Department, City College of the CUNY, New York, NY 10031}
\affiliation{CUNY Graduate Center, New York, NY 10031}
\author{J.~H.~Pixley}
\email{jed.pixley@physics.rutgers.edu}
\affiliation{Department of Physics and Astronomy, Center for Materials Theory, Rutgers University, Piscataway, NJ 08854 USA}
\author{Smitha Vishveshwara}
\email{smivish@illinois.edu}
\affiliation{Department of Physics, University of Illinois at Urbana-Champaign, Urbana, IL 61801-3080, USA}
\author{Bryce Gadway}
\email{bgadway@illinois.edu}
\affiliation{Department of Physics, University of Illinois at Urbana-Champaign, Urbana, IL 61801-3080, USA}
\date{\today}

\begin{abstract}
Using synthetic lattices of laser-coupled atomic momentum modes, we experimentally realize a recently proposed family of nearest-neighbor tight-binding models having quasiperiodic site energy modulation that host an exact mobility edge protected by a duality symmetry. These one-dimensional tight-binding models can be viewed as a generalization of the well-known Aubry-Andr\'{e} (AA) model, with an energy-dependent self duality condition that constitutes an analytical mobility edge relation. By adiabatically preparing the lowest and highest energy eigenstates of this model system and performing microscopic measurements of their participation ratio, we track the evolution of the mobility edge as the energy-dependent density of states is modified by the model's tuning parameter. Our results show strong deviations from single-particle predictions, consistent with attractive interactions causing both enhanced localization of the lowest energy state due to self-trapping and inhibited localization of the highest energy state due to screening.
This study paves the way for quantitative studies of interaction effects on self duality induced mobility edges.
\end{abstract}
\maketitle

Disorder-induced localization of quantum mechanical wavefunctions 
represents a fundamental change in the nature of eigenstates~\cite{Anderson-1958}. 
While electron-electron and electron-phonon interactions prohibit direct detection of single-particle localization in electronic systems, analog realizations of such phenomena have been made in coherent and controllable platforms based on photonic materials~\cite{Segev-Light-Loc} and ultracold atoms~\cite{SP2010}.
Some of the earliest observations of localization for both light~\cite{Lahini-AA-Light} and atoms~\cite{Roati08,Fallani-BoseGlass} were achieved in one dimension using a deterministic quasiperiodic potential in the Aubry-Andr\'{e} (AA) model~\cite{Aubry80,Thouless-AA,Sokoloff85,Holthaus-AA-Atoms-1997}.
However, the AA model is rather fine tuned and does not manifest a mobility edge, which separates localized states from extended ones as a function of energy.
Mobility edges, \textit{i.e.}, energy-dependent localization transitions, are expected to be the generic behavior of more general quasiperiodic models in one~\cite{Soukolis82, Liu85, thoulessprl88, sankarprl88, sankarprb90, Biddle-T1T2, Biddle-ExpT} and higher dimensions~\cite{lemarie2009universality,devakul,pixley2018weyl,Schneider-QC1,Schneider-QC2}. Mobility edges also accompany the appearance of delocalized states for models with short-range random 
disorder in higher dimensions~\cite{Evers-2008}.

\begin{figure}[t!]
	\includegraphics[width=0.95\columnwidth]{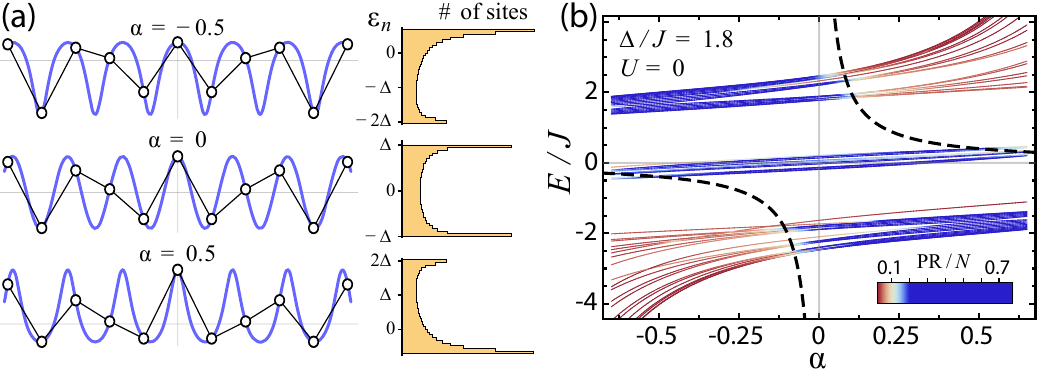}
	\caption{\label{FIG:fig1}
		\textbf{The generalized self-dual Aubry-Andr\'{e} model.}
		\textbf{(a)}~The generalized Aubry-Andr\'{e} potential and lattice site energies of Eq.~\eqref{EQ:GAA} shown for $\phi=0$ and tuning parameter $\alpha=-0.5,0,0.5$, with corresponding distributions of lattice site energies $\varepsilon_n$.
		\textbf{(b)}~Calculated eigenenergies and participation ratios (PR, in color) vs. $\alpha$ for a non-interacting model just below the critical quasiperiodicity strength at $\Delta/J=1.8$ ($N=51$ sites). Away from $\alpha=0$, eigenstates localize at different energies, forming a mobility edge.
		Dashed black lines show analytically predicted energy values of the ME (Eq.~\eqref{EQ:GAA_ME}).
}
\end{figure}

In recent years, mobility edges (MEs) in non-interacting models have been observed in three-dimensional disordered systems~\cite{jendrzejewski2012three,mcgehee2013three,LENS-ME,Lopez-KR}, as well as in reduced dimensions with quasiperiodicity in experiments based on ultracold atoms~\cite{Luschen-Bloch-SPME,An-MobEdge}. These observations were achieved using speckle disorder, bichromatic optical lattice experiments~\cite{Luschen-Bloch-SPME,Monika-Mob} harnessing native beyond-nearest-neighbor tunneling terms~\cite{Holthaus-naturalSPME,Li-SPME-Generic}, and in synthetic lattice experiments~\cite{An-MobEdge} based on the direct, independent engineering of next-nearest-neighbor tunneling~\cite{Biddle-ExpT}.
In these cases, however, accurate experimental control over the location of the mobility edge is lacking, as its analytic functional form is unknown. It is in principle possible to circumvent this issue in quasiperiodic systems by exploiting tight-binding models that have an \textit{exact} mobility edge that can be derived from an energy dependent self-duality condition~\cite{thoulessprl88,sankarprl88,sankarprb90,Biddle-ExpT,Ganeshan-Pixley-GAA}.
Experimental realization of an analytical mobility edge can help resolve the effects of interactions on the energy dependent localization transition, which remain a subtle and open theoretical question.

In this work, we experimentally realize a generalized \aub (GAA) model that has an exact mobility edge~\cite{Ganeshan-Pixley-GAA} and demonstrate control over the ME physics by employing synthetic lattices of laser-coupled atomic momentum modes~\cite{Gadway-KSPACE,Meier-AtomOptics}. Crucially, in the absence of interactions this model has an energy dependent self-duality that gives rise to the mobility edge.
In experiment, we probe the presence of the ME by measuring the localization properties of the lowest and highest energy states of the system, and vary the energy of the ME via a tuning parameter.
We map out comprehensive localization phase diagrams for these energy states, demonstrating that the ME is shifted by atomic interactions due to screening and self-trapping effects, but that overall the localization transitions and the ME survive.
Consequently, our work showcases the capacity of cold atomic setups for
the exploration of localization in quasiperiodic lattice models,
pinpointing the presence of a ME and capturing the important role of interactions.

The Hamiltonian realized in this work, $H_{tot} = H_{GAA} + H_{int}$, involves both the tight-binding GAA model proposed in Ref.~\cite{Ganeshan-Pixley-GAA} and a contribution due to atomic interactions. The GAA Hamiltonian is given by
\begin{equation}
H_{GAA} = -J \sum_{n} \left( c^\dagger_{n+1} c_{n} + \text{h.c.} \right) + \sum_{n} \varepsilon_n c^\dagger_n c_n,
\label{EQ:TBHam}
\end{equation}
where $J$ is a (real and positive) nearest-neighbor tunneling amplitude, $c_n$ destroys a boson at site (momentum mode) $n$,
and the GAA quasiperiodic site energies read
\begin{equation}
\varepsilon_n = \Delta \frac{\cos(2\pi n b + \phi)}{1 - \alpha \cos(2\pi n b + \phi)} ,
\label{EQ:GAA}
\end{equation}
with the quasiperiodicity amplitude and phase given by $\Delta$ and $\phi$, respectively. We choose the periodicity to be $b=\left(\sqrt{5}-1\right)/2$, though the localization results we present here hold for any irrational number~\cite{Ganeshan-Pixley-GAA}.
The tuning parameter $\alpha\in (-1,1)$ controls the shape of the potential and the resulting distribution of site energies, as shown by the blue curves in Fig.~\pref{FIG:fig1}{a}.
At $\alpha=0$, Eq.~(\ref{EQ:GAA}) reduces to the standard AA form, with a cosine dispersion and a cosine distribution of site energies, leading to an energy-independent localization transition.
For  $\alpha \neq 0$, the GAA model exhibits an exact 
ME at energy  $E$ 
following the relationship~\cite{Ganeshan-Pixley-GAA}
\begin{equation}
\alpha E= \, 2J-\Delta,
\label{EQ:GAA_ME}
\end{equation}
for the positive values of $J$ and $\Delta$ that we consider.

Atomic interactions further enrich the physics of this system. Low-energy, $s$-wave collisions between atoms in the various momentum modes~\cite{Ozeri-RMP} are described by
\begin{equation}
H_{int} = (U/2N_\textrm{at}) \sum_{i,j,k,l} c_i^\dagger c_j^\dagger c_k c_l \ .
\label{EQ:IntHam}
\end{equation}
Here
$U = g\rho$ is the mean-field interaction energy per atom for a sample of $N_\textrm{at}$ atoms occupying a single momentum mode,
$\rho$ is the atomic number density, $g = 4 \pi \hbar^2 a /M$ is the interaction term, $M$ is the atomic mass, and $a$ is the scattering length.
Collisions primarily conserve individual mode populations~\cite{SuppMats}, so we make the simplifying assumption of only considering mode-conserving collisions in our theoretical treatment.
Because the typical occupied site is populated by thousands of atoms,
we further treat the interactions through a mean-field Gross-Pitaevskii formalism.
The quantum statistics of the bosonic atoms leads to a strong mode-dependence of the pairwise interactions, 
and at the mean-field level they can be effectively described in terms of a purely local \textit{intra-mode attraction} with a collective energy scale $U$.
While our theoretical treatment ignores some details~\cite{SuppMats},
it provides a simple mean-field-level comparison that captures most of the salient features.

\begin{figure*}[t!]
\centering
	\includegraphics[width=.83\textwidth]{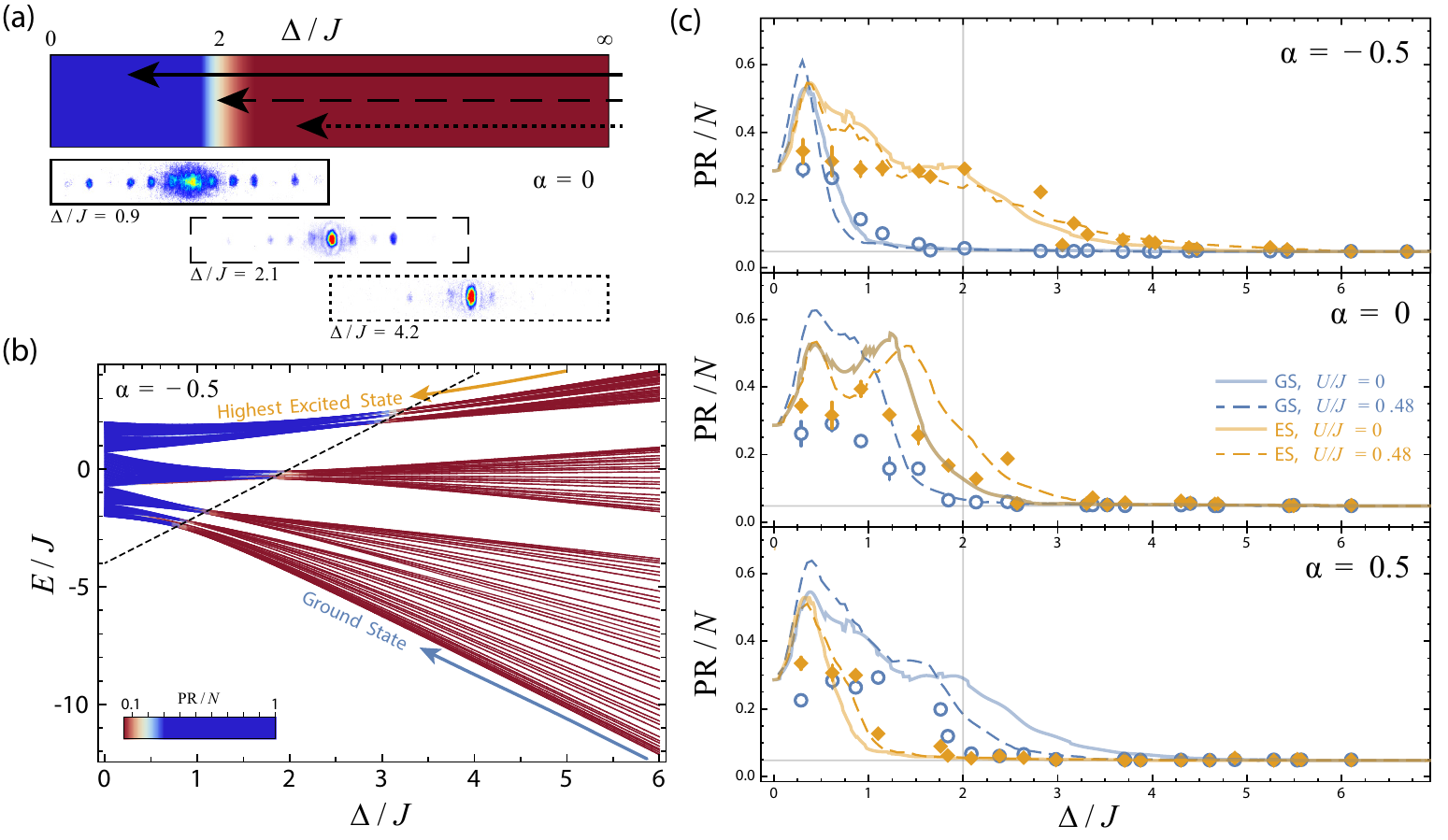}
	\caption{\label{FIG:fig2}
		\textbf{Probing the localization transition by adiabatic Hamiltonian evolution.}
		\textbf{(a)}~Cartoon of the experimental sequence (arrows). Population initially localized for $\Delta/J = \infty$ is slowly loaded into an eigenstate of the GAA model at a final quasiperiodicity-to-tunneling ratio $\Delta/J$.
        Bottom: Momentum distributions, corresponding to populations in the synthetic lattice, of the highest excited state for $\alpha = 0$ in the localized regime ($\Delta/J = 4.2$), near the delocalization transition  ($\Delta/J = 2.1$), and in the delocalized regime  ($\Delta/J = 0.9$).
		\textbf{(b)}~Numerically-calculated participation ratios (PR) overlaid on the eigenenergies of the GAA model for $\alpha=-0.5$, $\phi = \pi$, and $N = 201$~sites.
		High-energy states localize at larger quasiperiodicity strengths than low-energy states, highlighting the presence of the mobility edge of Eq.~\eqref{EQ:GAA_ME} (dashed black line).
		\textbf{(c)}~PR$/N$ vs. $\Delta/J$ for the ground (open blue circles) and highest excited states (yellow diamonds) under $\alpha={-0.5,0,0.5}$, showing evidence for a ME tunable via $\alpha$.
		Numerical Gross-Pitaevskii results include the exact experimental tunneling ramp and assume a homogeneous mean-field energy $U/J = 0.48$ ($U/h = 300$~Hz) for the dashed curves and zero interactions ($U/J = 0$) for the solid curves.
}
\end{figure*}

To probe the expected ME of this system, we determine the localization properties of the GAA eigenstates. We quantify localization through the participation ratio, $\textrm{PR} = 1 / \Sigma_n P_n^2$,
where $P_n$ is the normalized atom population at site $n$.
The PR effectively counts the number of sites that ``participate'' in hosting a state. It takes values ranging from $\textrm{PR} \sim N$ in the extended regime (\textit{e.g.},
$2N/3$ for the ground state of an $N$-site lattice with open boundaries) to $\textrm{PR} = 1$ for states localized to a single site.
For $\alpha \neq 0$, states on opposite sides of the ME correspond to PRs close to opposite extremes of this range, as depicted by the clear change in color in Fig.~\pref{FIG:fig1}{b}.

The strong dependence of localization behavior on $\alpha$ can be understood by considering how this parameter influences the distribution of site energies (see
Fig.~\pref{FIG:fig1}{a}).
For $\alpha < 0$, the effective site-energy potential is weighted towards higher energy values. In a heuristic picture, more sites ``sit'' on top of the wells rather than at their bottoms. 
Thus, for negative $\alpha$, a higher (lower) quasiperiodicity strength is required to induce localization for states at high (low) energy, as there are many more (fewer) nearby sites to which they can resonantly hop.
For positive values of $\alpha$, the complete opposite behavior is found, with the localization behavior of the high and low energy states swapped.
In this way the ME is directly controllable through the parameter $\alpha$, as suggested by Eq.~(\ref{EQ:GAA_ME}).

We experimentally realize the GAA model with control over the parameter $\alpha$ in a synthetic lattice~\cite{OzawaPrice-SynthDim} of coupled atomic momentum modes~\cite{Gadway-KSPACE,Meier-AtomOptics}.
We start with an optically trapped Bose--Einstein condensate of $\sim 10^5$ $^{87}$Rb atoms, with the atoms having nearly zero momentum.
We then use a pair of counter-propagating lasers (wavelength $\lambda=1064$~nm) to drive two photon Bragg transitions, based on virtual absorption from and stimulated emission into the applied laser beams, allowing for the atoms to change their momentum in increments of $2\hbar k$ (with $k=2\pi/\lambda$ and $\hbar$ the reduced Planck's constant).
While one of the two lasers has a single frequency tone, the other beam is engineered to have many distinct components.
The individual frequency components combine with the single-frequency beam to address a unique two-photon Bragg transition, thus creating an effective ``tunneling link'' between the synthetic lattice ``sites'' (relating physically to modes with momentum values $p_n = 2n\hbar k$, with $n$ the site index).
By independently tuning the strength, phase, and detuning from Bragg resonance of each of these terms, we respectively control the tunneling amplitude, tunneling phase, and site-to-site energy difference of each link in the corresponding synthetic lattice.
In this work, we make use of this generic site energy control to exactly implement the GAA potential of Eq.~\eqref{EQ:GAA} on a $21$-site lattice for $|\alpha| \leq 0.5$~\cite{SuppMats}.

To explore the presence of a ME, we seek to adiabatically prepare the lowest and highest energy eigenstates of the system.
We initialize population in the central site
of a lattice with all tunneling links set to $0$ and with GAA site energies imposed. The phase term of Eq.~(\ref{EQ:GAA}) is set to be $\phi=\pi$ ($0$) to ensure that the initial lattice site has the lowest (highest) energy.
We slowly ramp up the tunneling from $0$ to a final value of $J/h = 625$~Hz over $0.75$~ms, and hold at that value for $1.25$~ms.
At the single-particle level and in its adiabatic limit, this ramping procedure prepares the lowest (highest) energy eigenstate of the full Hamiltonian
when initializing at the lowest (highest) energy site in the zero-tunneling limit~\cite{SuppMats}.
Our ramp can alternatively be viewed as tuning the system from the limit of infinite quasiperiodicity ($\Delta/J = \infty$, where our initialized state maps to a strictly localized eigenstate), to a final quasiperiodicity-to-tunneling ratio $\Delta/J$, as shown in Figs.~\pref{FIG:fig2}{a,b}.
By repeating this procedure for different combinations of $\Delta$ and $\alpha$, we map out the localization behavior of the extremal eigenstates of the GAA model across a broad range of parameters.

We expect this procedure to be robust in the insulating regime, where there is poor overlap and weak coupling between the system's localized eigenstates.
However, the finite ramp duration will lead to diabatic corrections, particularly important as the eigenstates hybridize upon encountering a delocalization transition.
Thus, while this procedure may not fully capture eigenstate properties in the metallic regime, we expect that it is well-suited for determining the delocalization transition
for a given eigenstate and $\alpha$ value.

Figure~\pref{FIG:fig2}{a} demonstrates this procedure performed for the highest energy state of the canonical AA model ($\alpha=0$), demonstrating localization above the critical quasiperiodicity strength $\left(\Delta/J\right)_c=2$ and extended delocalization below it.
By studying the localization properties of the lowest and highest energy eigenstates (ground state denoted ``GS'' and highest excited state denoted ``ES''), we expect to find evidence of an energy-dependent localization transition
when $\alpha\neq 0$.
Concretely, the numerically-calculated PR values of the eigenstates in the non-interacting limit for $\alpha=-0.5$ are shown in Fig.~\pref{FIG:fig2}{b}. They illustrate a clear energy dependence in agreement with the prediction of Eq.~\eqref{EQ:GAA_ME} (dashed black line), with the GS and ES localization transitions found near $\Delta/J=1$ and $\Delta/J=3$, respectively.

The experiment, however, features atomic interactions that can shift the localization transitions away from single-particle predictions.
We capture this
numerically by 
solving the Gross-Pitaevskii equation (GPE) for a homogeneous mean-field interaction energy of $U/h = 300$~Hz ($U/J = 0.48$ in terms of the final tunneling value)~\cite{SuppMats}. Interacting GPE simulations of the PR values are shown in Fig.~\pref{FIG:fig2}{c} as the dashed blue (yellow) lines for the GS (ES), taking account the exact parameter ramp used in experiment. For comparison, simulation results for $U = 0$ are shown as shaded solid lines. 

Figure~\pref{FIG:fig2}{c, top} shows the energy-dependent localization behavior for $\alpha = -0.5$. We plot the normalized PR values, PR/$N$, which should range from $1/21$ (gray horizontal line) in the site-localized limit to $\lesssim 2/3$ in the extended regime.
Roughly speaking, we observe PR/$N$ values that remain low for a range of large $\Delta/J$ values, which then increase as the states undergo delocalization transitions.
From the distinct separation of the localization transitions for the GS and ES we can infer the existence of an intervening ME.

Consistent with the GPE simulations, we do not observe a significant influence of interactions for $\alpha=-0.5$.
The $\alpha=0$ case reduces to the standard AA model. Thus, in the absence of interactions, all eigenstates should delocalize at the same critical value of $\Delta/J = 2$.
However, as shown in Fig.~\pref{FIG:fig2}{c, center}, we observe that the transition in fact splits for the lowest and highest energy states, signaling a mobility edge that arises solely from atomic interactions~\cite{Schreiber842}.
For $\alpha=+0.5$ (Fig.~\pref{FIG:fig2}{c, bottom}), our data show an inversion of the mobility edge: the excited state localizes at a weaker quasiperiodicity amplitude than the ground state.
This inversion is expected due to a symmetry of the non-interacting Hamiltonian ($H_{GAA}$) that exchanges the lowest and the highest energy states as $\alpha \to -\alpha$ (and $\phi \to \phi + \pi$ for an exact inversion in a finite system).
We additionally observe a strong shift of the GS delocalization transition away from the non-interacting theory prediction for $\alpha=+0.5$.

We find qualitative agreement with the behavior expected based on the GAA model, observing a ME that inverts as we go from $\alpha = -0.5$ to $\alpha=+0.5$.
However, we do not observe the simple symmetry between the GS and ES predicted by the GAA model (Eq.~\ref{EQ:GAA}) as $\alpha$ is taken from negative to positive values. Instead, we find the asymmetric response detailed above, with a larger magnitude of separation between the observed GS and ES transitions for $\alpha=+0.5$ as compared to $\alpha=-0.5$, and the appearance of a mobility edge even for the $\alpha = 0$ case.
These observations are consistent with interaction-driven shifts of the transitions and the fact that the interacting GAA model has an enlarged symmetry, by which the GS and ES localization properties exchange if we take $U \rightarrow -U$ as $\alpha \rightarrow -\alpha$.
These results demonstrate that, despite interactions strongly breaking the self-dual symmetry of the non-interacting model, the ME is renormalized and survives many-body effects.

Our simple mean-field description of the system's effectively local and attractive interactions~\cite{An-Inter,Ozeri-RMP,SuppMats} allows us to provide an intuitive picture for how the localization properties of the GS and ES are respectively affected. 
For states at low energy, the chemical potential shifts due to interactions inhibit delocalization in the synthetic lattice. This instability towards self-trapping for attractive interactions~\cite{Giamarchi-Loc} shifts the ground state localization transition towards lower quasiperiodicity strengths for all values of $\alpha$.
In contrast, for states at high energy, attractive interactions can effectively screen the GAA quasiperiodic potential, promoting delocalization. 
This screening by attractive interactions for high energy states is analogous to the more familiar screening by local repulsive interactions for bosonic ground states~\cite{Giamarchi-Loc,Deissler-DisorderWithInteractions-2010}.

\begin{figure}[t!]
	\includegraphics[width=0.77\columnwidth]{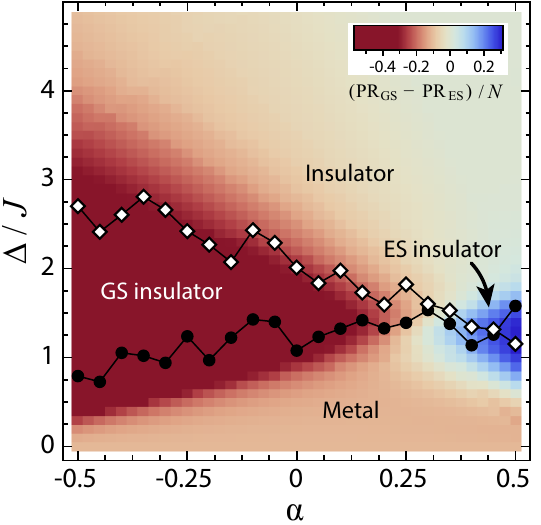}
	\caption{\label{FIG:fig3}
		\textbf{Localization phase diagram of the GS and ES.}
		Critical quasiperiodicity values for the onset of GS and ES delocalization (filled circles and open diamonds), overlaid on the difference in normalized participation ratio (PR/$N$, with difference shown according to the inset colorbar) of the numerically calculated extremal eigenstates for a mean-field interaction $U = 0.48 J$.
		The GS and ES transition ``lines'' do not coincide, indicating a mobility edge, and they cross away from $\alpha=0$, indicating a shift due to atomic interactions.
	}
\end{figure}

Figure~\ref{FIG:fig3} provides a more comprehensive picture for the localization behavior of the GAA model with interactions, achieved by studying the GS and ES localization transitions for a larger set of $\alpha$ values. For the GS and ES, we perform the same state preparation ramps as described for Fig.~\ref{FIG:fig2}, starting from the $\Delta/J = \infty$ limit. For each sampled $\alpha$ value, we determine the ``critical'' $\Delta/J$ at which delocalization occurs, relating to an increase of the normalized participation ratio (PR/$N$) above a threshold value set to $0.19$. The collections of critical $\Delta/J$ values, shown respectively as white diamonds and black disks for the ES and GS, serve to define the boundaries for the onset of delocalization for these states. 

In the absence of interactions, these two curves should be symmetric about an inversion of $\alpha \to -\alpha$, with a crossing at $\alpha = 0$ that relates to the absence of a ME in the canonical AA model. However, we observe that interactions lead to a significant deviation from this non-interacting expectation.
The crossing of these two localization transition lines is clearly shifted away from $\alpha = 0$, appearing at $\sim 0.3-0.4$. This behavior is in agreement with the expectations from the interaction phenomena of self-trapping and screening.

Beneath the data, we show the numerically calculated (by imaginary time propagation) difference in PR/$N$ for the GS and ES for a homogeneous interaction energy $U = 0.48 J$. This calculated difference of the participation ratios reveals a behavior that is similar to what is observed from the experimental data. The theory plot exhibits a shift of the crossing point away from $\alpha = 0$. It also indicates a region at large $\Delta/J$ in which both states are insulating, and a region at small $\Delta/J$ in which both states are metallic. Finally, it shows two regions in which a mobility edge can be directly inferred based on the localization of only one of these states.

Together, the experimental transition lines and the simulation results can be viewed as the localization phase diagram for the extremal states of the GAA model with local, attractive mean-field interactions. The system exhibits interaction shifts to its localization transitions, as well as a parameter-tunable ME that survives the effects of interactions.
Because the extremal energy states are the first or final states to undergo a localization transitions for increasing quasiperiodicity amplitude, the combined upper and lower boundaries in Fig.~\ref{FIG:fig3} can be viewed as defining the critical boundaries for the onset of a mobility edge, as eigenstates begin to localize (delocalize) for increasing (decreasing) quasiperiodicity.

Here, we've presented the first experimental realization of an exact mobility edge by emulating the generalized Aubry-Andr\'{e} model in the presence of interactions~\cite{Ganeshan-Pixley-GAA}. We mapped out the localization phase diagram of the lowest- and highest-energy states of the system and found evidence for a parameter tunable mobility edge. We observed shifts to the localization transitions due to interaction effects, relating to self-trapping and screening for the low and high energy states, respectively. In the future, by combining with injection-based spectroscopy techniques~\cite{Cheuk-injection}, these results may be extended to allow the precise determination of the energy of the mobility edge in this and other quasiperiodic models~\cite{Weld-Phasonic}, as well as to determine the role of critical wavefunctions in greatly enhancing interaction effects~\cite{PhysRevLett.98.027001,PhysRevLett.109.246801,PhysRevLett.108.017002,PhysRevB.89.155140,PhysRevB.101.235121}.

\section{Acknowledgments}
This material is based upon work  supported by the Air Force Office of Scientific Research under Grant No.~FA9550-18-1-0082 (F.A.A., E.J.M., and B.G.) and  through Grant No.~FA9550-20-1-0136 (J.H.P). S.G. was supported by NSF OMA-1936351. K.P. and S.V. acknowledge support by NASA (SUB JPL 1553869).

\bibliographystyle{apsrev4-1}
\bibliography{GAA-JP-SG}

\end{document}